 \documentstyle[prl,aps,psfig,amssymb,multicol]{revtex}

\begin{document}
\draft
\title{Spectroscopy with random and displaced random ensembles}
\author{ V. Vel\'azquez and A. P. Zuker}
\address{ IRES, B\^at27, IN2P3-CNRS/Universit\'e Louis
Pasteur BP 28, F-67037 Strasbourg Cedex 2, France}
\date{\today}
\maketitle
\begin{abstract}
  Due to the time reversal invariance of the angular momentum operator
  $J^2$, the average energies and variances at fixed $J$ for random
  two-body Hamiltonians exhibit odd-even-$J$ staggering, that may be
  especially strong for $J=0$. It is shown that upon ensemble
  averaging over random runs, this behaviour is reflected in the yrast
  states. Displaced (attractive) random ensembles lead to rotational
  spectra with strongly enhanced $BE2$ transitions for a certain class
  of model spaces. It is explained how to generalize these results to
  other forms of collectivity.
\end{abstract}
\begin{multicols}{2}
\narrowtext 
  
Even-even nuclei are more bound than their odd neighbours (OES,
odd-even staggering). Their ground states have always spin $J=0$ (J0D,
$J=0$ dominance), and with excedingly few exceptions their first
excited state has $J=2$ and decays through an enhanced $BE2$
transition. These systematic features have been traditionally
interpreted in terms of specific components of the nuclear
interaction.  OES, for instance was attributed to the pairing force
but it has been argued that, in small Fermi systems, it is a universal
phenomenon associated with deformation~\cite{hak97}.  Though the claim
is too strong, there is some truth in it as it can be viewed as a
consequence of the Jahn-Teller effect which produces OES through the
filling of time-reversed pairs in doubly degenerate single particle
states, e.  g., Nilsson orbits~\cite{sat98}.
  
The discovery that random interactions lead to $J=0$ ground and $J=2$
first excited states far more often than statistically
expected~\cite{joh98} has brought the subject of systematic features
into sharper focus. In particular, OES is often present, even in the
absence of either pairing forces or deformed fields~\cite{joh99},
which means that it should not be associated to a specific form of
coherence. We are left to find general causes. 

In what follows it will be argued that time reversal (${\cal T}$)
invariance is at the origin of OES and J0D. Later, it will be shown
that the attractive nature of the interaction leads to $BE2$ coherence
so far detected in a randomized IBM context~\cite{bij00} but not in
shell model (SM) simulations. 

Before going into the simulations themselves, we define notations and
collect some basic results and predictions based on ${\cal T}$
invariance.  

We shall rely for insight on the low moments of the Hamiltonian
$H${\em at fixed} $J$, which determine the level
densities~\cite{can01}.  Calling $d_J$ the dimensionality of the
vector space, the centroid is $E_{cJ}=d_J^{-1}\text{tr}(H)_J\equiv
d^{-1}_J\langle H \rangle_J$, and the variance
$\sigma^2_J=d_J^{-1}\langle (H-E_{cJ})^2\rangle_J$. 

Using $r,s,\ldots$ for subshells and $i\equiv rr_z\ldots$ for
individual orbits, $H$ can be written in uncoupled or coupled form, in
terms of pair creation and annihilation operators
\[H=\sum_{i<j,k<l}W_{ijkl}Z^+_{ij}Z_{kl}=
\sum_{r<s,t<u,J}W_{rstu}^JZ^+_{ij,J}\cdot Z_{kl,J}.\]

The $W_{rstu}^J$ matrix elements will be taken to belong to the
Gaussian orthogonal ensemble (GOE), i.e., to be real and normally
distributed with mean zero and variance $v^2$ for the off-diagonals
and $2v^2$ for the diagonals. For $n>2$ the Hamiltonian matrices are
said to belong to the two body random ensemble (TBRE).  An alternative
should be considered: GUE, where U stands for unitary, in which case
the matrix elements are complex (general Hermitian, rather than
symmetric, matrices). TBRE derived from GOE Hamiltonians are ${\cal
  T}$-invariant, while those derived from GUE ones are
not~\cite{bro81}.

In Ref.~\cite{bij99} it was found that both ensembles lead to the same
behaviour for OES and J0D. At first sight the result is surprising
since ${\cal T}$ invariance is at the origin of such behavior, as we
shall see soon. The paradox is resolved by noting that the $J^2$
operator is ${\cal T}$ invariant, and states of good $J$, transform as
${\cal T}|JM\langle=(-)^{J+M}|J-M\rangle$~\cite{BM69}. At $M=0$, which
contains the information on all states, eigenstates of $J^2$ are also
eigenstates of ${\cal T}$, with eigenvalue $(-)^J$: true for TBRE
derived from either GOE or GUE. Now examine the consequences.
  
In a determinantal basis, at $M=0$, if a state and its time
reversed are different, the resulting even and odd combinations
(e.g., $|jm\, j'-m\rangle\pm (-)^{j+j'}|j'm\, j-m\rangle$) contribute
equally to the traces of $H^k$. The self conjugate states are always
even (e.g., $|jm\, j-m\rangle$). Therefore, at fixed number of
particles $n$ there are always more even $J$ states than odd ones.

Obviously, the difference in average energies between even and odd
states, $\zeta$, will depend on the sign for the average over
selconjugate states. Upon ensemble averaging $\zeta$ will vanish, and
{\em we expect no staggering between $E_{cJ}$ even and odd states}. It
is equally obvious that for the variances all contributions are
positive. Hence {\em the average $\sigma^2_J$ will always be larger
  for even states than for odd ones, and staggering will persist after
  ensemble averaging}.

For odd $n$, ${\cal T}$ has no consequences because it has no
eigenstates; since ${\cal T}^2=-1$, the $(1\pm {\cal T})|JM\rangle$
combinations change sign under ${\cal T}$.

To (try to) explain J0D we note that all the $J=0$ states of seniority
zero are in the self-conjugate space. It is fairly simple to calculate
their centroids and widths. If they contribute substantially to the
totals, much of the $J$-OES will be due to them and they will
concentrate on the overall $J=0$ averages thus explaining their
special status. A strong hint in favor of this argument is the massive
presence of seniority zero states indicated by the persistence of
pair-transfer coherence in $J=0$ ground states~\cite{joh99}.
To sum up all the indications related to ${\cal T}$: 

Whatever separates even and odd $J$ is due to the selfconjugate space,
and whatever is special about $J=0$ is due to the seniority zero
states, entirely contained in this space. And, remember, ${\cal T}$
invariance does not {\em imply} J0D, it only {\em suggests} its
frequent observation.
\begin{figure}[htb]  
    \begin{center}
      \leavevmode 
      \psfig{file=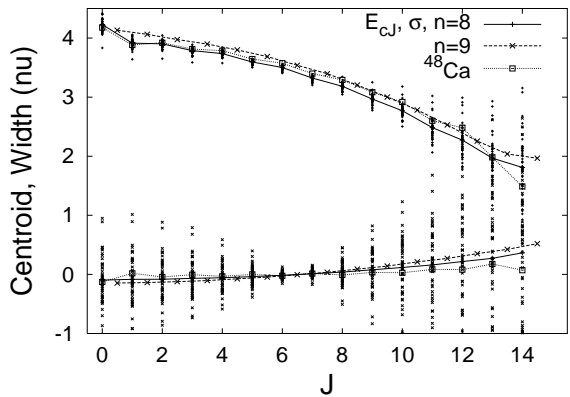,width=8.cm} 
      \caption{$(pf)^{8,9}_J$ centroids and widths. See text }
      \label{fig:1}
    \end{center}
\end{figure}

Now do 50 runs for $n=8,9$ in the $pf$ shell, normalizing each to unit
scalar variance $\sigma^2(n=2)=1$ . (For the calculation of traces we
refer to~\cite{gi73,ja81,ve82}, though here we use a home-made
method.)

In Fig.~\ref{fig:1} the dots show $E_{cJ}$ (below) and $\sigma_J$
(above) for the 50 $n=8$ runs, while the---full, dashed and
dotted---lines are the averages for $n=8,9$, and the results for the
full KB3 interaction plus single particle field in
$^{48}$Ca~\cite{cau94}---respectively. 

As expected, the $E_{cJ}$ ensemble average is smooth and for
$\sigma_J$ the OES between spins ($J$-OES) for $n=8$ persists after
ensemble averaging while $\sigma(9)_J$ does not stagger; and though
larger on the average than $\sigma(8)_J$, its maximum is below
$\sigma(8)_{J=0}$.

The random runs were done with monopole-free interactions ($\sum_J
W_{rsrs}^J(2J+1)=0$), which simplify considerably the calculation of
$\sigma_J$ with negligible loss of accuracy. The $^{48}$Ca results are
exact (all eigenvalues for all $J$ were calculated~\cite{can01}). Note
that---to within a trivial overall factor---KB3 with its carefully
tuned monopoles and experimental single particle spacings yields the
same pattern for $\sigma_J$ as the monopole-free ensemble average.  No
assumptions on two body monopoles will be made in what follows.

The next task is to {\em prove} that $E_{cJ}$ and $\sigma_J$ are
sufficient to determine the {\em ensemble averaged} yrast patterns.
The proof is as follows: The low moments of $H$ determine a smooth
tridiagonal matrix. Once diagonalized it leads to a smooth binomial
that describes very well the level densities. The position of the
yrast state depends on the parameter $N_J=\ln d_J/ln 2$. If the third
moment vanishes---as assumed here---the energy converges to the lower
bound $E_{bJ}=E_{cJ}-\sqrt{N_J\sigma^2_J}$ ~\cite{can01}. For moderate
values of $N_J$ ($\approx 10$ in our example) this limit is missed by
$\approx 2\sigma_J/\sqrt{N_J}$ and a good estimate demands explicit
diagonalization, which will obviously reflect $J$-OES.

This result is strictly true for {\em ensemble averages}. For
individual matrices, because of fluctuations, the bound is only an
unreliable estimate (an old story under a new
guise~\cite{bro81,rat71}). For example: with KB3, in $^{48}$Ca the
$J\ne 0$ yrast states come below their exact positions by 0.25 to 2.5
MeV, while the ground state comes 4 MeV too high.

The subject demands a full treatement, but the arguments above are
sufficient to confirm the crucial role of $\sigma_J$ anticipated in
\cite{bij99} but ignored in~\cite{mul00}, and put in doubt
in~\cite{kuz00} through a misunderstanding that deserves a comment:
Boson simulations are conducted in a collective subspace of much
smaller dimensions than the corresponding fermion problem. In
particular, the $N_J\times N_J$ tridiagonals in ~\cite{kuz00} (once
corrected~\cite{bij01}) are interesting models for the Lanczos
submatrix at the origin~\cite{can01}, but it makes no sense to compare
their variance to that of the full $d_J\times d_J$ matrix
($d_J=2^{N_J}$).

\vspace{.2cm}

Of the three general properties mentioned in the first paragraph,
$BE2$ enhancement is the one not spontaneously produced by purely
random trials. Some coherence is needed, and the only way to simulate
it, in a GOE context, is through a constant displacement, $c$, of all
matrix elements. The displaced GOE (DGOE) is a standard ensemble, for
which there exists a famous result: its spectrum is a semicircle, with
one detached level~\cite{bro81}.  Therefore, for attractive forces,
the coherent part is a matrix whose elements are constants $c=-|c|$;
leading to a displaced TBRE (DTBRE).

In a two-body context with good $J$, attractive forces do not have
systematically negative matrix elements, but their signs must have a
very general origin, because all realistic interactions are
spectacularly similar~\cite{duf96}, and the {\em extremely} rare sign
discrepancies only affect the smallest absolute values.  Furthermore,
as was noted in Ref.~\cite{cor82}, these signs are strongly correlated
to those of Elliott's $q\cdot q$ force. Hence, there may be a---DTBRE
{\em vs} $q\cdot q$---``sign coherence'' conflict.

Quite conveniently, it is possible to analize---and then resolve---it,
while staying within the DTBRE; by examining results in two spaces.
The first is a ($\Delta j=2$) subspace of a major shell, in which all
$q\cdot q$ and realistic matrix elements are negative. It consists of
the orbits with $j=l+1/2$, for which a quasi-$SU3$ symmetry can
operate, leading to quadrupole properties similar to those of the full
shell~\cite{zuk95}.  The second is the major shell itself.

Accordingly, two sets of runs were performed. The first in the $\Delta
j=2$, $(f_{7/2}p_{3/2})$ space ($fp$ for short) respects sign
coherence. The other, in the full $pf$ shell, does not. {\em Please do
not confuse} $fp$ and $pf$. 

960 $fp$ runs were done for each of four combinations of number of
particles and isospin: $nT=84\, (^{48}$Ca), 40 ($^{44}$Ti), 61
($^{46}$Ti), and 80 ($^{48}$Cr) , with strict GOE interactions plus
single particle splitting. The parameters were chosen to mock
realistic values ($v=0.6,\, \epsilon_{f_{p3/2}}-\epsilon_{f_{7/2}}=2$
MeV). The $c=-1$ steps for DTBRE are arbitrary.

\begin{figure}[htb]  
  \begin{center}
    \leavevmode 
    \psfig{file=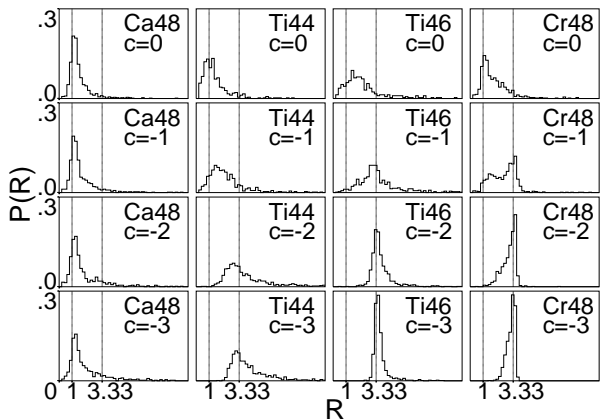,width=8.cm,angle=270}
    \vspace{.5cm}   
      \caption{Probability density for energy ratios $E_4/E_2$}
      \label{fig:4}
    \end{center}
\end{figure}

Fig.~\ref{fig:4} shows the evolution as a function of $c$ of the
$R=E_4/E_2$ energy ratio, a time-honored indicator of collective
behaviour: $R=1$, 2, 3.33 corresponding to seniority, vibrational and
rotational regimes respectively.
\vspace{-.5cm} 

\begin{figure}[htb]  
  \begin{center}
    \leavevmode 
    \psfig{file=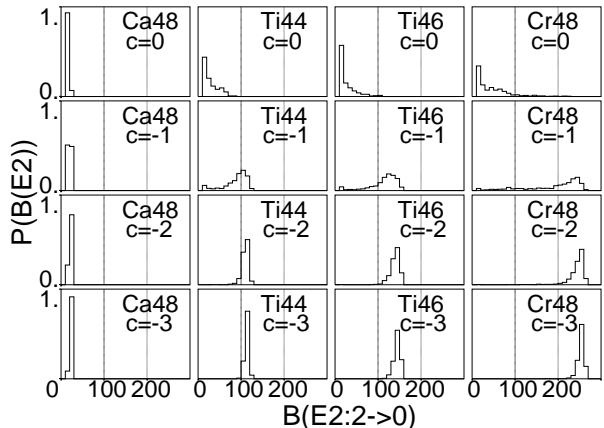,width=8.cm,angle=270}
    \vspace{.5cm}  
    \caption{Probability densities for $BE2(2\rightarrow 0)$   }
    \label{fig:5}
  \end{center}
\end{figure}

Only runs with $J=0$ ground states were kept and the probability
density $P$ normalized to unity for the four nuclei. At
$c=0$, the peaks are centered at $R=1$ except for $^{46}$Ti. As $c$
decreases $^{48}$Ca will not budge, while for its partners $R=3.33$
will be either clearly or overwhelmingly favored.  There is more than
a hint of rotational behaviour here, as confirmed by the distributions
of $BE2(2\rightarrow 0)$ strength in Fig.~\ref{fig:5}. It is seen that
at $c=0$ there is no coherence: the probabilities are consistent with
a Porter-Thomas law.  At $c=-1$ the rates increase sharply, and at
$c=-2,\, -3$ they concentrate in narrow peaks, whose strength should
be compared to that of a rotor of intrinsic quadrupole moment $Q_0$,
and $BE2(2\rightarrow 0)=0.02A^{2/3}\, Q_0^2\, e^2 fm^4$.

Following Ref.~\cite{zuk95}, we calculate the maximum
$Q_0$ values for $^{44}$Ti, $^{46}$Ti, and $^{48}$Cr in the $fp$ space
to be (22, 25, 32) (a-dimensional oscilator values) leading to
transitions of (120, 160, 270) $e^2fm^4$ respectively.  The main peaks
are seen to be very close to these values, confirming the rotational
nature of the $T=0,$ and 1 spectra.

To check the influence of sign coherence, some 200 runs for $^{48}$Cr
were performed in the full $pf$ shell, with uniform single particle
spacings of 2 MeV (as seen in $^{41}$Ca). The similarities with the
sign-coherent $fp$ trends are as significant as are the differences.
In both cases, as $c$ becomes more negative there is a consistent
buildup of $BE2$ strength, and a consistent increase of $J=0$ ground
states (see Table~\ref{tab:c} for the $fp$ runs), increasingly
associated to perfect yrast sequences (i.e, J=0,2,4,6,...).

%\vspace{-.3cm} 

The discrepancies can be summed up by saying that in the full $pf$ shell
nothing is as clearcut as in the $fp$ space, especially the limiting
behavior for large displacements.

The situation at $c=-2$ is typical.  In the $pf$ runs we find ``only''
75\% of $J=0$ ground states (against 99.8\% in $fp$), and practically
as many perfect sequences. The indicators are good---but more spread
than in Figs.~\ref{fig:4} and~\ref{fig:5}---at $E_4/E_2=2.92\pm 0.56$
and $BE2(2\rightarrow 0)=276\pm 23\, e^2 fm^4$ (the $SU3$ limit is 320
$e^2fm^4$, the strength is equally large in $J\ne 0$ ground states).
More disturbingly, the runs do not seem to converge to perfect rotors:
$c=-3$ brings little change over $c=-2$.  The fairly good rotational
behaviour suggested by the averages above is not systematically
confirmed by a well defined intrinsic state. In other words: $Q_0$ is
not always approximately constant for the lowest members of the band.
In the $fp$ case, at $c=-2$, 99.8\% of the sequences are perfect and
all yrast levels are those of a rotor. 

Statistically, sign-coherence is not indispensable to generate
acceptable $BE2$ enhancements, but physically it matters: Even nuclei
do not have $J=0$ ground states 50 or 80\% of the time, but 100\% of
the time.
\begin{table}[b]
\begin{center}
\begin{tabular}{cccccc}
$c$                   &  & $^{48}$Ca & $^{44}$Ti&$^{46}$Ti&$^{48}$Cr \\
\hline
$\ \ 0.0$                 &   & 76.3   & 46.6    & 33.2   & 60.2 \\
$-1.0$                    & & 72.8   & 59.4    & 65.0   & 88.1 \\
$-2.0$                    & & 61.1   & 82.0    & 94.9   & 99.8 \\
$-3.0$                    & & 53.3   & 94.7    & 99.7   & 100.0 \\
\end{tabular}
\caption{ Percentage of $J=0$ ground states as a function of
displacement $c$.} 
\label{tab:c}
\end{center}
\end{table}

$^{48}$Cr has better than 75\% chances of having a rotational-like
behaviour: it {\em is} a backbending rotor.

\begin{figure}[htb]  
  \begin{center}
    \leavevmode 
    \psfig{file=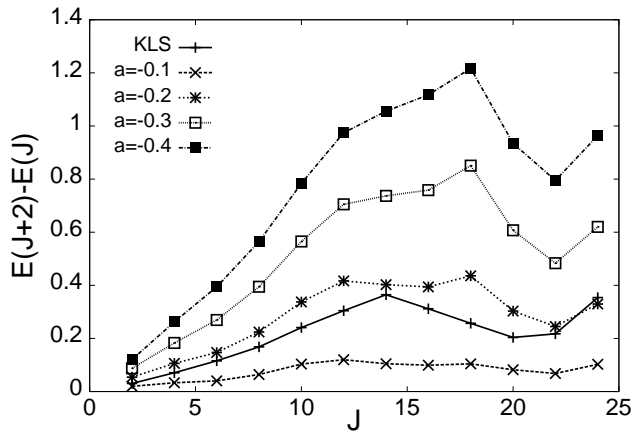,width=8.cm,angle=270}
\vspace{.3cm}
\caption{Backbending patterns in $(gds)^8$ $T=0$, with KLS interaction
    from~\protect\cite{zuk95}, and constant $W_{rstu}^{JT}=-a$.} 
    \label{fig:6}
  \end{center}
\end{figure}

Fig.~\ref{fig:6} illustrates what a constant interaction
$W_{rstu}^{JT}=-a$ does in the $\Delta j=2, \, (gds)^8\, T=0$ space.
With single particle spacings of 1 MeV, the $E_{J+2}-E_{J}$ patterns
are those of backbending rotors. They show more structure than the
realistic interaction results~\cite{zuk95} but the physics is very
much the same.

\vspace{-.3cm}

At this point, as far as DTBRE in the $pf$ shell goes, we are where we
we were twenty years ago in the $sd$ shell: The $BE2$ coherence is
there but ``The spectra are neither rotational nor particularly
interesting''~\cite{cor82}. But now we have the $fp$ spectra:
rotational, and therefore interesting. The difference is obviously due
to sign-coherence. But: Why do we have sign-coherence in $fp$? {\em
  Because in $\Delta j=2$ spaces, the signs, by construction, are the
  same as in $LS$ coupling}~\cite{zuk95}.

The heart of the problem is that we do, routinely, simulations in $jj$
coupling because it is the one used in shell model codes, as commanded
by the central---more generally monopole---field. However, the
``residual'' two body multipole part of the Hamiltonian is dominated
by forces that are overwhelmingly central (quadrupole, pairing,
etc.)~\cite{duf96}.  For these, the natural coupling scheme is $LS$.
Therefore, it is artificial to define for them DTBRE coherence through
the $jj$ matrix elements.

The solution of the sign-coherence problem becomes evident: define the
DTBRE in $LS$ scheme. Technically, we can continue to employ $jj$
scheme using the $LS$ displacement calculated once and for all. It is
a safe bet---backed by the $sd$ experience~\cite{cor82}---that the
$pf$ results will become as satisfactory as the $fp$ ones.
 
Our examples involve rotational motion, but now it should be possible
to generate other forms of collectivity, as seen by the simplest
example: with the same two body interaction the $(sd)^4$ spectrum is
rotational ($^{20}$Ne) and the $(sd)^{-4}$ spectrum is vibrational
($^{36}$Ar). The difference between the two is entirely due to the
change of central field. Therefore, to obtain both rotors and
vibrators it seems convenient to randomize the single particle
energies and fix the two body terms, contrary to what is usually done.

\vspace{.2cm}

We have identified time reversal invariance as the origin of odd-even
staggering of the mass surfaces and $J=0$ spin for the ground states
of even nuclei. The attractive nature of the forces appears to
provide a sufficient condition $BE2(2\rightarrow 0)$ enhancements
associated to collective behaviour. 

\vspace{-.5cm}

\begin{acknowledgements}
  The exact diagonalizations were performed with the ANTOINE code. The
  authors thank E. Caurier, J. Hirsch, F. Nowacki, and S. Rombouts for
  their help and comments.

 V.V. is a fellow of Conacyt (M\'exico).
\end{acknowledgements}

\end{multicols}
\end{document}